\documentclass[final,5p,times,twocolumn]{elsarticle}

\usepackage{amssymb}
\usepackage{amsmath}

\journal{Computer Methods and Programs in Biomedicine}

\usepackage{amsmath}
\usepackage{amssymb}
\usepackage{booktabs}
\usepackage{multirow}
\usepackage{array}
\usepackage{graphicx}
\usepackage{xcolor}
\usepackage{url}
\usepackage[hidelinks]{hyperref}

\begin{document}

\begin{frontmatter}



\title{MorphoBranch: A Fine-Structure-Preserving Workbench for
Morphometric Analysis of Branched Cellular Structures}




\author[aff1]{Zhiying Song}
\ead{zhiysong@zufe.edu.cn}

\author[aff2]{Hanyi Ling}
\ead{hanyiling@zju.edu.cn}

\author[aff3]{Junyi Wu}
\ead{junyiwu@zufe.edu.cn}

\author[aff4]{Nenggan Zheng}
\ead{zng@cs.zju.edu.cn}

\author[aff3]{Yangbo Jiang\corref{cor1}}
\ead{jiangyangbo@zufe.edu.cn}

\cortext[cor1]{Corresponding author}

\affiliation[aff1]{
    organization={School of Data Science, Zhejiang University of Finance and Economics},
    city={Hangzhou},
    postcode={310018},
    country={China}
}

\affiliation[aff2]{
    organization={College of Biomedical Engineering \& Instrument Science, Zhejiang University},
    city={Hangzhou},
    postcode={310027},
    country={China}
}


\affiliation[aff3]{
    organization={School of Information Technology and Artificial Intelligence,
    Zhejiang University of Finance and Economics},
    city={Hangzhou},
    postcode={310018},
    country={China}
}

\affiliation[aff4]{
    organization={Qiushi Academy for Advanced Studies,
    Zhejiang University},
    city={Hangzhou},
    postcode={310027},
    country={China}
}

\begin{abstract}
\textbf{Background and Objectives:}
Fluorescence-labeled cellular arbors provide readouts of neuronal and
microglial morphology, but fine and weakly labeled processes are prone to fragmentation and false connections that bias skeleton-based measurements. We present MorphoBranch, a fine-structure-preserving, human-reviewable workbench for morphometry of branched cellular structures.

\textbf{Methods:}
MorphoBranch combines a deterministic Morphometry Engine with an
LLM-assisted Refinement Engine. The Morphometry Engine implements an image-to-graph workflow integrating
multiscale structural evidence extraction, hysteresis segmentation,
evidence-constrained skeleton refinement, and graph-based morphometry. The Refinement
Engine maps natural-language requests to registered actions for parameter
adjustment, preview execution, metric reporting, and unsupported-request
handling, while image processing and quantitative computation remain
deterministic and reviewable.

\textbf{Results:}
MorphoBranch was evaluated on two public neuronal axon datasets, AxonMIP
and AxonStack, and the in-house CellMorph dataset of microglial fluorescence images. It achieved the highest Skeleton F$_1$ and clDice and the lowest
length-estimation error among the evaluated methods on all three datasets,
while also achieving the highest Dice and IoU on AxonMIP and AxonStack.
Across 150 natural-language tasks, the Refinement Engine achieved a 94.0\%
end-to-end success rate.


\textbf{Conclusions:}
These results demonstrate that MorphoBranch provides a reproducible, human-reviewable workflow for
morphometric analysis of branched cellular structures. It supports
fine-structure-preserving quantification across neuronal axon and microglial 
fluorescence images while maintaining inspectable and reproducible
analysis workflows.
\end{abstract}




\end{frontmatter}



\section{Introduction}


Many neurons and ramified microglia possess elaborate cellular arbors
composed of slender processes and complex branching patterns. In neurons,
axonal and dendritic branching contributes to circuit connectivity and signal
integration, whereas microglial processes dynamically extend and retract to
survey the surrounding tissue \cite{Cuntz2010,Nimmerjahn2005}. Morphometric
properties such as process length, branching complexity, and spatial extent
therefore provide quantitative structural readouts of cellular architecture
and remodeling \cite{Hansen2022,Kim2024}. Accurate and reproducible
quantification of these properties is consequently important for
characterizing meaningful structural variation across cells and diverse
biological conditions.


Obtaining such measurements reliably remains challenging. A typical
skeleton-based workflow segments labeled cellular structures from a
fluorescence $z$-stack or its two-dimensional projection, converts the
resulting mask into a skeleton graph, and derives morphometric measurements
from this representation. Because these stages are sequentially coupled,
the analysis faces two related problems. First,
\textit{\textbf{fine-structure ambiguity}} arises because cellular processes are often
thin, weakly labeled, and densely distributed. Uneven background, imaging
noise, signal attenuation, crossing, and overlap can make faint processes
difficult to distinguish from artifacts and obscure their connectivity,
while projection may further superimpose structures from different depths.
Second, \textit{\textbf{topological error propagation}} occurs when missed processes,
artificial breaks, or false connections introduced during segmentation are
propagated through skeletonization into spurious components, endpoints,
junctions, branches, and paths, thereby biasing measurements of process
length, branching complexity, and radial organization. Addressing these
problems requires repeated ROI selection, parameter adjustment, and
manual inspection, making the analysis time-consuming, operator-dependent,
and difficult to reproduce across samples and experiments.

A variety of software tools have been developed for quantitative morphology
analysis, supporting structure tracing, reconstruction, visualization, and
morphometric analysis~\cite{arshadi2021snt,peng2014vaa3d}, as well as
image-based segmentation and quantification of morphological features
~\cite{narro2007neuronmetrics,dehmelt2011neuritequant,
ho2011neurphologyj}. Specialized tools further address particular
structures or cell types, including neuronal branch terminals
~\cite{kanaoka2019determ} and microglial morphology~\cite{Hansen2022}.
Despite these capabilities, two task-specific limitations remain for
morphometric analysis of fine, weakly labeled, and densely branched cellular
structures in fluorescence images. First, \textit{\textbf{cross-stage
integration and measurement-level validation}} remain incomplete. Reliable
morphometry requires fine-structure preservation and topology control to be
coordinated throughout image-to-graph analysis and evaluated through
downstream morphometric measurements rather than segmentation alone. Second,
\textit{\textbf{iterative refinement and workflow traceability}} remain
challenging. Dataset-dependent adjustments often span multiple interfaces
and custom scripts, increasing operator burden while leaving parameter
changes, quality-control outputs, manual decisions, and software versions
incompletely documented. These limitations motivate an integrated,
human-reviewable system that combines fine-structure-preserving morphometry
with efficient and traceable refinement.

To address these, we present \textbf{MorphoBranch}, a task-oriented
workbench for fine-structure-preserving morphometric analysis of branched
cellular structures in fluorescence images. MorphoBranch centers on a
deterministic \textbf{Morphometry Engine} and optionally provides an \textbf{LLM-assisted
Refinement Engine}, together with functions for data management,
quality control, provenance tracking, and result export. Specifically, the
Morphometry Engine integrates fine-structure-preserving structure extraction
with topology-aware skeleton and graph analysis to improve the reliability
of downstream morphometric measurements. In addition, building on recent
advances in natural-language scientific analysis
~\cite{ye2023amadeusgpt,zhong2026neuroagent}, the Refinement Engine
translates user requests into structured and editable analysis plans. Users
can preview, revise, and confirm adjustments, reducing dependence
on manual operations and custom scripting while retaining human
oversight. Approved plans are executed by deterministic analysis modules,
with configurations and execution records retained as provenance,
thereby separating flexible LLM-assisted interaction from reproducible
quantitative computation. Finally, we evaluate MorphoBranch using public
neuronal axon datasets, in-house fluorescence microscopy images of
microglia, and a natural-language interaction benchmark to assess structural
fidelity, morphometric accuracy, computational efficiency, and refinement
reliability.
Our contributions are summarized as follows:
\begin{itemize}

\item We develop \textbf{MorphoBranch}, an integrated workbench with a deterministic,
fine-structure-preserving Morphometry Engine that combines structure
extraction with topology-aware skeleton and graph analysis for reliable
morphometric analysis of fine, weakly labeled and densely branched cellular
structures in fluorescence images.

\item We introduce an \textbf{LLM-assisted, human-in-the-loop Refinement Engine} that
translates natural-language requests into editable analysis plans and
supports preview, revision, and confirmation. Deterministic execution and
provenance tracking make the refined analyses reviewable and reproducible
across iterative refinements.

\item We conduct system-level evaluations on public neuronal axon datasets,
the in-house CellMorph microglial dataset, and a \textbf{
natural-language interaction benchmark} (150 tasks), demonstrating the end-to-end utility of MorphoBranch
for reliable morphometry and controlled workflow refinement across
complementary settings.
\end{itemize}

\begin{figure*}[t]
    \centering
    \includegraphics[width=0.9\textwidth]{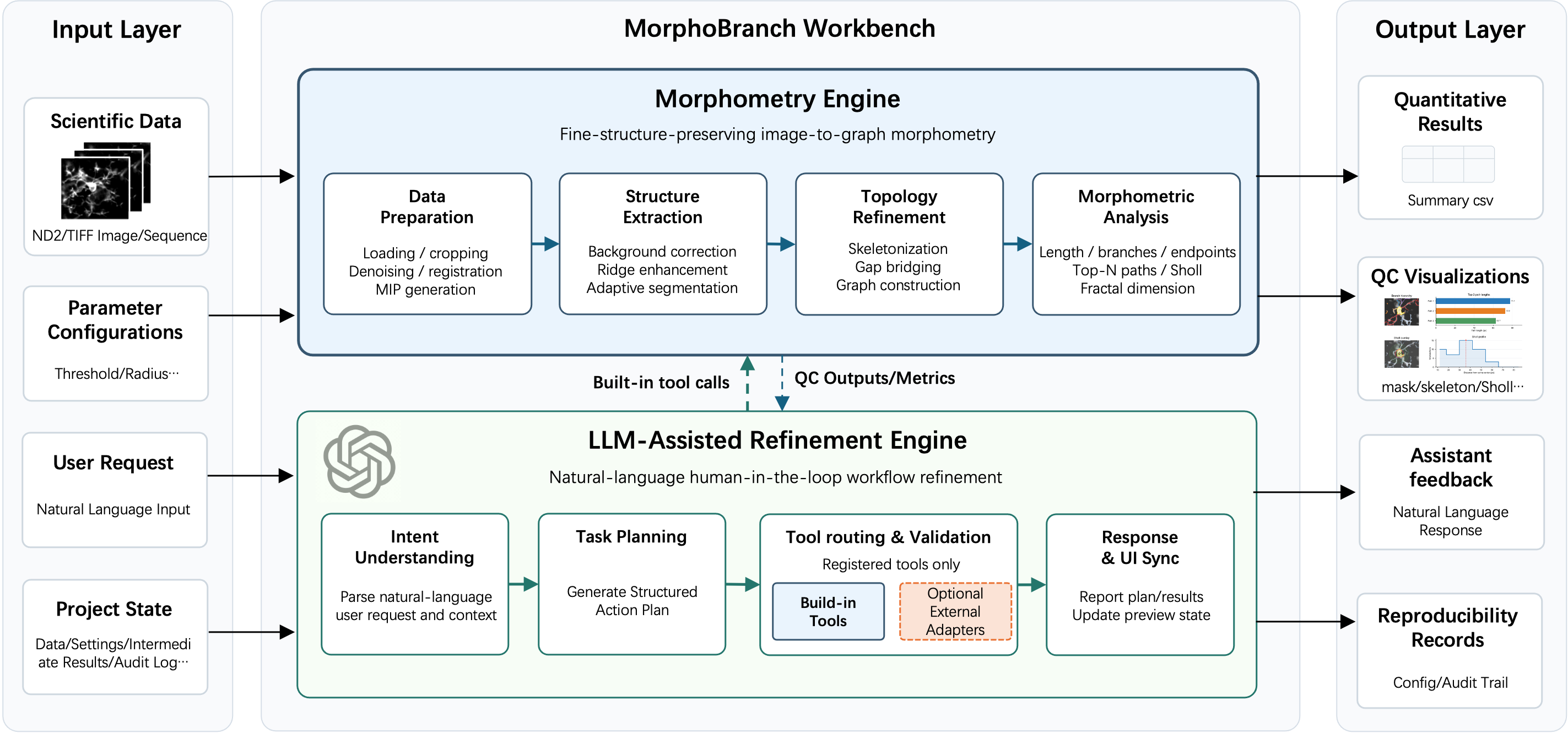}
    \caption{Overall architecture of MorphoBranch. The workbench integrates a
deterministic Morphometry Engine with an optional LLM-assisted Refinement
Engine. The Morphometry Engine performs data preparation, structure
extraction, topology refinement, and morphometric analysis, whereas the
Refinement Engine translates natural-language requests into validated
workflow adjustments through registered actions. Quantitative results,
quality-control visualizations, refinement feedback, and reproducibility
records are exported for review and downstream analysis.}
    \label{fig:architecture}
\end{figure*}

\section{Method}
\subsection{System Overview and Design Principles}
MorphoBranch is designed to preserve structural and topological fidelity
throughout image-to-graph morphometric analysis while enabling efficient,
human-reviewable, and traceable dataset-specific refinement. As shown in
Fig.~\ref{fig:architecture}, the system comprises a deterministic Morphometry Engine, an optional LLM-assisted Refinement Engine, and supporting modules for data and ROI management, quality control, workflow-state and provenance tracking, and result export.
Its architecture follows three design principles:
\begin{itemize}
\item \textbf{Cross-stage deterministic morphometry.}
The Morphometry Engine coordinates fine-structure-preserving structure
extraction with topology-aware skeleton and graph analysis across processing
stages, limiting the loss of faint processes and the propagation of local
artifacts into morphometric measurements.

\item \textbf{Constrained and human-reviewable refinement.}
The Refinement Engine translates natural-language requests into structured configuration changes that are reviewed and confirmed before execution. This reduces reliance on repeated manual operations and custom scripting while retaining user oversight.

\item \textbf{Provenance-preserving reproducibility.}
Supporting modules retain project data, ROIs, configurations, quality-control outputs, user decisions, and execution records. Confirmed analyses can therefore be inspected and replayed without requiring the language model.
\end{itemize}

Operationally, the Morphometry Engine forms the primary computational
path from microscopy data to quantitative results. Structure extraction and
skeleton refinement are detailed in Figs.~\ref{fig:structure_extraction}
and~\ref{fig:skeleton_refinement}, respectively. The LLM-assisted
Refinement Engine provides optional interaction across configurable
stages of this deterministic workflow, as shown in
Fig.~\ref{fig:llm_refinement}. All confirmed adjustments are executed
by the same deterministic modules, separating flexible workflow control
from reproducible quantitative computation.

\subsection{Project and Workflow-State Management}
\label{sec:workflow_management}

MorphoBranch organizes each analysis as a project linking immutable ND2 or TIFF
inputs and their metadata to ROI definitions, spatial calibration, analysis
configurations, intermediate artifacts, quality-control outputs, and exported
results. Each execution creates a versioned state whose artifacts remain
associated with their source image, configuration, and parent state. A
confirmed state represents the active accepted analysis, whereas a preview
state contains provisional artifacts that cannot overwrite raw inputs or
confirmed outputs. When a configuration value changes, a dependency manager
invalidates the affected stage and its downstream artifacts while preserving
compatible upstream results. This state model prevents incompatible artifacts
from being combined and supports efficient preview, restoration, comparison,
and deterministic rerunning.

\subsection{Deterministic Morphometry Engine}
\label{sec:morphometry_engine}
The Morphometry Engine is the deterministic analytical core of MorphoBranch.
It addresses fine-structure ambiguity and topological error propagation
through an evidence-coupled image-to-graph workflow. Rather than relying
only on a binary mask after segmentation, the engine retains both a
structure mask and a continuous structural evidence map, allowing skeleton
refinement to be constrained by image support.
As shown in Fig.~\ref{fig:architecture}, the engine comprises four
components: Data Preparation, Structure Extraction, Topology Refinement,
and Morphometric Analysis. Data Preparation generates analysis
images from raw microscopy inputs; Structure Extraction produces the
binary structure mask and structural evidence map; Topology Refinement
repairs supported discontinuities and rejects unsupported connections; and
Morphometric Analysis computes geometric, topological, branch-hierarchy, and 
spatial-complexity descriptors. Intermediate outputs are retained for
quality-control inspection, provenance tracking, and deterministic reruns.

\subsubsection{Data Preparation}
\label{sec:data_preparation}

Data Preparation converts two-dimensional images or ND2/TIFF microscopy
stacks into prepared analysis images. For image stacks, users specify
the target channel, ROI, and $z$-range, and the selected stack region is
projected into a two-dimensional image using a configured projection
operator \(\mathcal{P}\), with maximum-intensity projection used by default:
\begin{equation}
I(x,y)=
\mathcal{P}\left(\{I_0(x,y,z):z\in\mathcal{Z}\}\right),
\end{equation}
where \(I_0\) is the input stack and \(\mathcal{Z}\) is the selected
slice range. Precomputed two-dimensional images can be used directly
without projection. Optional registration can be applied to image
sequences to reduce frame-to-frame displacement. 
The resulting analysis image \(I(x,y)\), together with the ROI, channel and
projection settings when applicable, and spatial calibration when available,
is passed to Structure Extraction.

\begin{figure}[t]
    \centering
    \includegraphics[width=\linewidth]{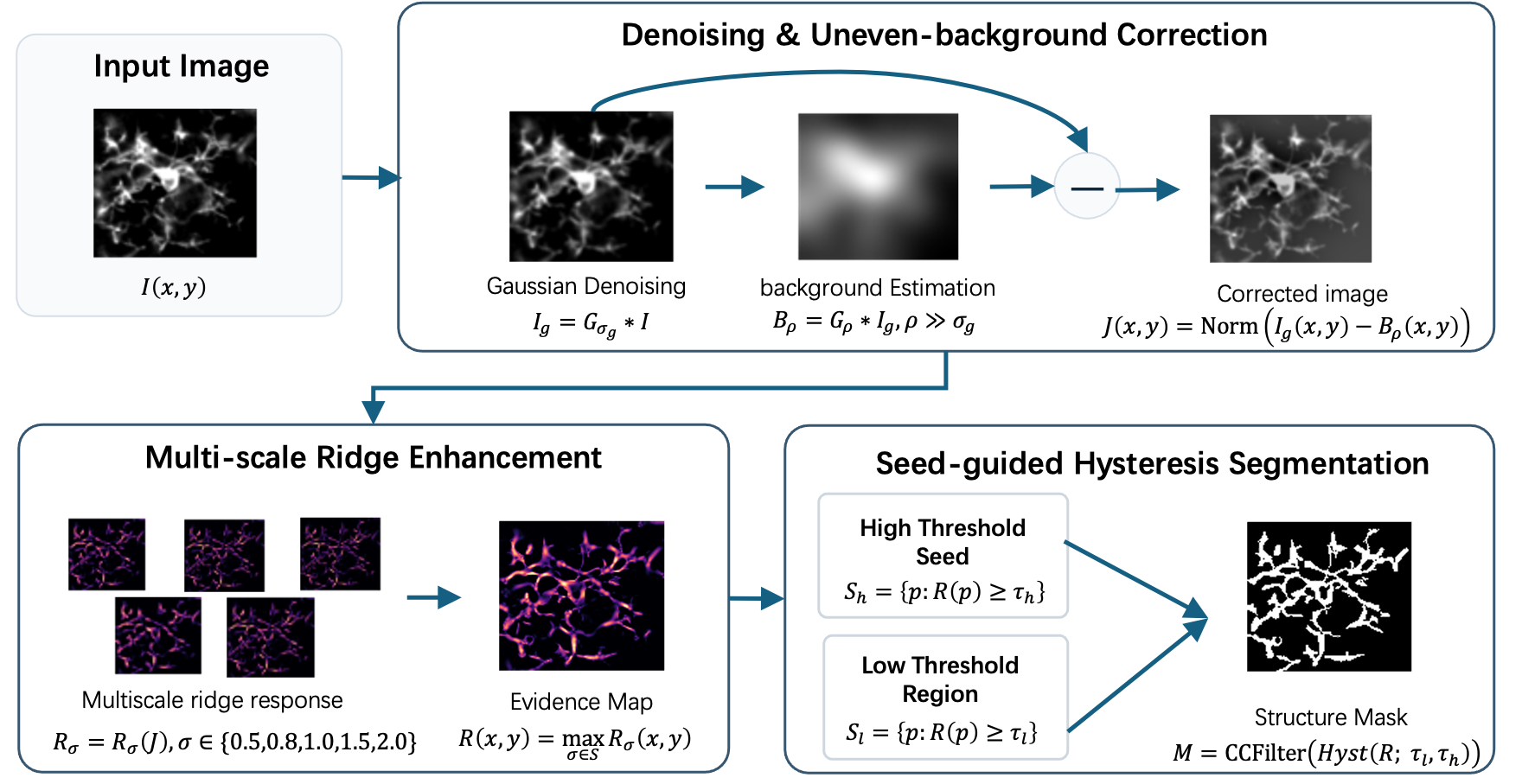}
    \caption{Structure extraction workflow. The prepared analysis image
    $I(x,y)$ is denoised, corrected for slowly varying background, and
    transformed into a multiscale structural evidence map $R$. Hysteresis
    thresholding identifies high-confidence seed regions and connected
    low-threshold structures, producing a binary structure mask while
    preserving weak, continuous processes.}
    \label{fig:structure_extraction}
\end{figure}

\subsubsection{Structure Extraction}
\label{sec:structure_extraction_method}
Structure Extraction addresses the ambiguity between faint cellular
processes and imaging artifacts by producing two complementary
representations: a binary structure mask $M$ and a continuous structural
evidence map $R$ (as shown in Fig.~\ref{fig:structure_extraction}). Given the prepared
analysis image $I(x,y)$, MorphoBranch first applies Gaussian denoising and
broad-scale background subtraction to reduce high-frequency noise and
slowly varying background illumination:
\begin{equation}
I_g = G_{\sigma_g} * I,\qquad
B_{\rho}=G_{\rho}*I_g,\quad \rho\gg\sigma_g,
\end{equation}
\begin{equation}
J=\operatorname{Norm}\left(\max(I_g-B_{\rho},0)\right),
\end{equation}
where $G_{\sigma_g}$ suppresses high-frequency noise, $B_{\rho}$
estimates slowly varying background illumination, and $J$ denotes the
normalized background-corrected image.

Multiscale Meijering filtering~\cite{meijering2004design} enhances
curvilinear structures of different apparent widths:
\begin{equation}
R(x,y)=
\max_{\sigma\in\mathcal{S}}
\mathcal{R}_{\sigma}(J)(x,y),
\end{equation}
where $\mathcal{R}_{\sigma}$ denotes the ridge response at scale
$\sigma$, and $\mathcal{S}$ is the configured scale set. The binary
structure mask is then obtained by seed-guided hysteresis
thresholding~\cite{canny1986computational}:
\begin{equation}
S_h=\{p:R(p)\ge\tau_h\},\qquad
S_l=\{p:R(p)\ge\tau_l\},
\end{equation}
\begin{equation}
M=\operatorname{RemoveSmall}
\left(
\operatorname{Hyst}(R;\tau_l,\tau_h)
\right),
\end{equation}
where $\operatorname{Hyst}(R;\tau_l,\tau_h)$ retains low-threshold
regions connected to high-threshold seeds, and
$\operatorname{RemoveSmall}(\cdot)$ removes connected components smaller
than a predefined minimum component size.
This procedure preserves
weak but connected processes while suppressing isolated background
responses. Both $M$ and $R$ are passed to Topology Refinement, allowing
candidate repairs to use continuous image evidence rather than the
binary mask alone.
When a visible soma or compact proximal structure is present, MorphoBranch
can also assist reference-point initialization from the corrected image
and structure mask for downstream morphometric analysis.

\begin{figure}[t]
    \centering
    \includegraphics[width=\linewidth]{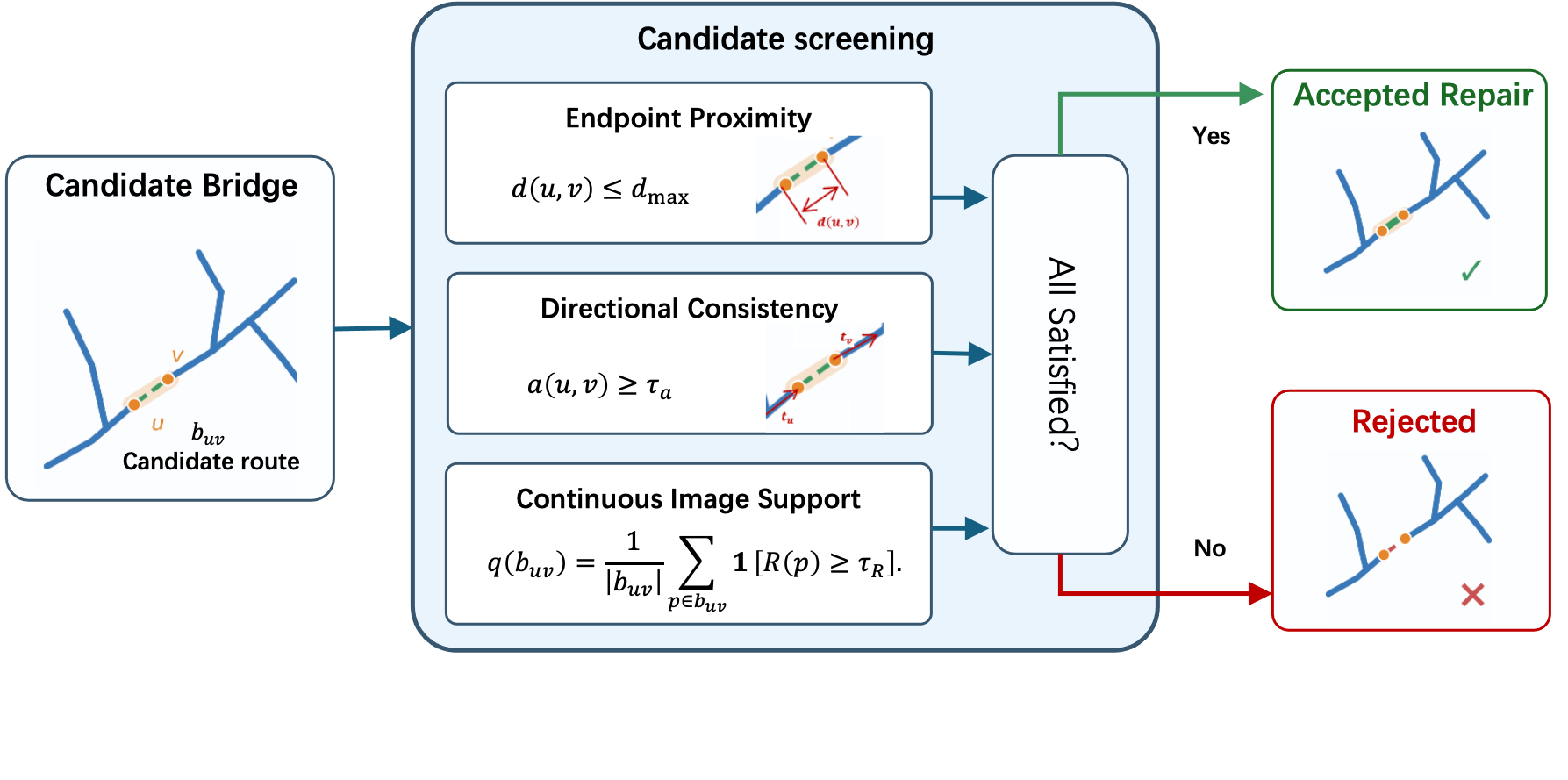}
    \caption{Evidence-constrained skeleton refinement. Candidate bridges
    between skeleton endpoints are screened using endpoint proximity,
    directional consistency, and continuous image evidence. A candidate
    repair is accepted only when all criteria are satisfied, allowing
    image-supported interruptions to be repaired while rejecting unsupported
    shortcuts.}
    \label{fig:skeleton_refinement}
\end{figure}

\begin{table*}[t]
\centering
\caption{Core morphometric descriptors reported by MorphoBranch,
organized into four categories characterizing geometry, topology,
branch hierarchy, and spatial complexity of the extracted structures.}
\label{tab:morphometric_descriptors}
\footnotesize

\resizebox{\linewidth}{!}{%
\begin{tabular}{@{}p{2.0cm}p{3.2cm}p{7.0cm}p{6.2cm}@{}}
\toprule
\textbf{Category} &
\textbf{Descriptor} &
\textbf{Definition} &
\textbf{Interpretation} \\
\midrule

\multirow{3}{*}{Geometry}
& Foreground area
& $A=|M|s_xs_y$, where $|M|$ is the number of foreground pixels
& Area occupied by the extracted structure \\

& Total skeleton length
& $L_{\mathrm{total}}=\sum_{e\in E}\ell(e)$, where $\ell(e)$ is the length of edge $e$
& Overall extent of the skeletonized processes \\

& Skeleton density
& $\rho=L_{\mathrm{total}}/A$
& Skeleton extent normalized by foreground area \\

\midrule

\multirow{6}{*}{Topology}
& Connected components
& Number of connected subgraphs in the refined graph $G^{*}$
& Structural continuity and potential fragmentation \\

& Endpoints
& $N_{\mathrm{end}}=|\{v\in V:\deg(v)=1\}|$
& Peripheral terminations and potential skeleton breaks \\

& Junctions
& $N_{\mathrm{junc}}=|\{v\in V:\deg(v)\geq3\}|$ after junction consolidation
& Branching complexity and potential false connections \\

& Graph branch segments
& Maximal paths between structural nodes via degree-2 nodes
& Number of distinct skeleton segments \\

& Branch-length distribution
& $\{\ell(b_i)\}_{i=1}^{N_b}$ for all graph branch segments $b_i$
& Distribution of local branch lengths \\

& Longest geodesic path
& $L_{\mathrm{geo}}=\max_{u,v\in V_{\mathrm{end}}}d_G(u,v)$
& Longest continuous trajectory in the graph \\

\midrule

\multirow{3}{*}{Branch hierarchy}
& Branch segments by order
& Counts of segments by branch order
& Reference-centered hierarchical branching \\

& Maximum branch order
& $O_{\max}=\max_{b_i}O(b_i)$, where $O(b_i)$ is the order of branch $b_i$
& Maximum depth of hierarchical branching \\

& Mean branch length by order
& $\bar{L}_{k}=N_k^{-1}\sum_{i:O(b_i)=k}\ell(b_i)$ for branch order $k$
& Branch lengths across hierarchical levels \\

\midrule

\multirow{4}{*}{Spatial complexity}
& Sholl intersection profile
& $N(r)$, skeleton intersections at radius $r$ from the reference point
& Radial distribution of branching complexity \\

& Critical radius
& $r_{\mathrm{c}}=\arg\max_r N(r)$
& Radius of maximal Sholl complexity \\

& Peak intersections
& $N_{\max}=\max_r N(r)$
& Maximum number of Sholl intersections \\

& Fractal dimension
& $D=\lim_{\varepsilon\to0}\frac{\log N_B(\varepsilon)}{\log(1/\varepsilon)}$, estimated by linear regression
& Scale-dependent spatial complexity \\

\bottomrule
\end{tabular}%
}
\end{table*}

\subsubsection{Topology Refinement}
\label{sec:topology_refinement_method}

Topology Refinement addresses the propagation of mask
discontinuities and pixel-level skeleton artifacts into spurious
components, endpoints, junctions, and paths. It converts the
structure mask $M$ and structural evidence map $R$ into a refined
skeleton graph (Fig.~\ref{fig:skeleton_refinement}). 
After limited morphological closing and one-pixel skeletonization, an initial graph
is constructed as
\begin{equation}
S_0=\operatorname{Skel}\!\left(\operatorname{Close}(M)\right),\qquad
G_0=\Gamma(S_0)=(V_0,E_0),
\end{equation}
where $\operatorname{Skel}(\cdot)$ denotes one-pixel-wide skeletonization,
and $\Gamma(\cdot)$ converts the skeleton into a graph whose nodes
correspond to endpoints and junction regions and whose edges represent
the intervening skeleton paths.

To reconnect likely true interruptions while avoiding false shortcuts
introduced by spatial proximity alone, candidate bridges are generated
between skeleton endpoints. For a candidate bridge $b_{uv}$ connecting
endpoints $u$ and $v$, MorphoBranch screens endpoint proximity,
directional consistency, and continuous image support as
\begin{equation}
\begin{aligned}
d(u,v) &= \|\mathbf{x}_u-\mathbf{x}_v\|_2 \le d_{\max},\\
a(u,v) &\ge \tau_a,\\
q(b_{uv}) &=
\frac{1}{|b_{uv}|}
\sum_{p\in b_{uv}}
\mathbb{I}\!\left[R(p)\ge\tau_R\right]
\ge \tau_q .
\end{aligned}
\end{equation}
Here, $\mathbf{x}_u$ and $\mathbf{x}_v$ denote the image coordinates of
endpoints $u$ and $v$, respectively, and $d(u,v)$ is their Euclidean
distance, constrained by the maximum bridging distance $d_{\max}$.
The term $a(u,v)$ measures the compatibility between the local endpoint
directions and the candidate bridge direction, with $\tau_a$ denoting
the corresponding acceptance threshold. For a candidate bridge $b_{uv}$,
$|b_{uv}|$ denotes its number of pixels, and $\mathbb{I}[\cdot]$ is the
indicator function. The quantity $q(b_{uv})$ is the fraction of bridge
pixels whose structural evidence exceeds $\tau_R$, and $\tau_q$ is the
minimum required support ratio. Only bridges satisfying all three
constraints are accepted, allowing image-supported discontinuities to be
reconnected while rejecting unsupported shortcuts caused by noise,
projection overlap, or segmentation artifacts. Adjacent junction pixels
are consolidated to avoid inflated junction counts, and edge lengths are
computed using spatial calibration when available before morphometric
analysis.


\subsubsection{Morphometric Analysis}
\label{sec:morphometric_analysis}

Morphometric Analysis converts the structure mask $M$ and refined
graph $G^{*}=(V,E)$ into quantitative descriptors, as summarized in
Table~\ref{tab:morphometric_descriptors}. These descriptors are grouped into
four categories: geometry, topology, branch hierarchy, and spatial
complexity. Geometry descriptors quantify structural coverage and skeleton
extent, including total skeleton length:
\begin{equation}
L_{\mathrm{total}}=\sum_{e\in E}\ell(e),
\end{equation}
where $\ell(e)$ denotes the length of graph edge $e$.
Topology descriptors characterize graph connectivity and path organization,
including connected components, endpoints, junctions, branch segments,
branch-length distributions, and longest geodesic path. Endpoints and
consolidated junctions define the structural nodes used to partition the
skeleton into graph branch segments.
For structures with a reference region, branch-hierarchy
descriptors organize graph segments into primary, secondary, and higher-order
branches and quantify branch-order-specific counts and lengths. Additional
reference-centered path summaries and branch visualizations are provided
when a reference region is available.
Spatial-complexity descriptors characterize radial organization and
scale-dependent complexity using Sholl analysis
\cite{sholl1953dendritic} and box-counting fractal dimension. The Sholl
reference point can be initialized from a detected reference region when
available or specified by the user when no soma or proximal anchor is
present, and is retained with the exported measurements.
Length- and area-based descriptors are reported in physical units when
spatial calibration is available and otherwise in pixel units. The
reference point used for branch-hierarchy and Sholl analyses is recorded
with the exported measurements.

\begin{figure}[t]
    \centering
    \includegraphics[width=\linewidth]{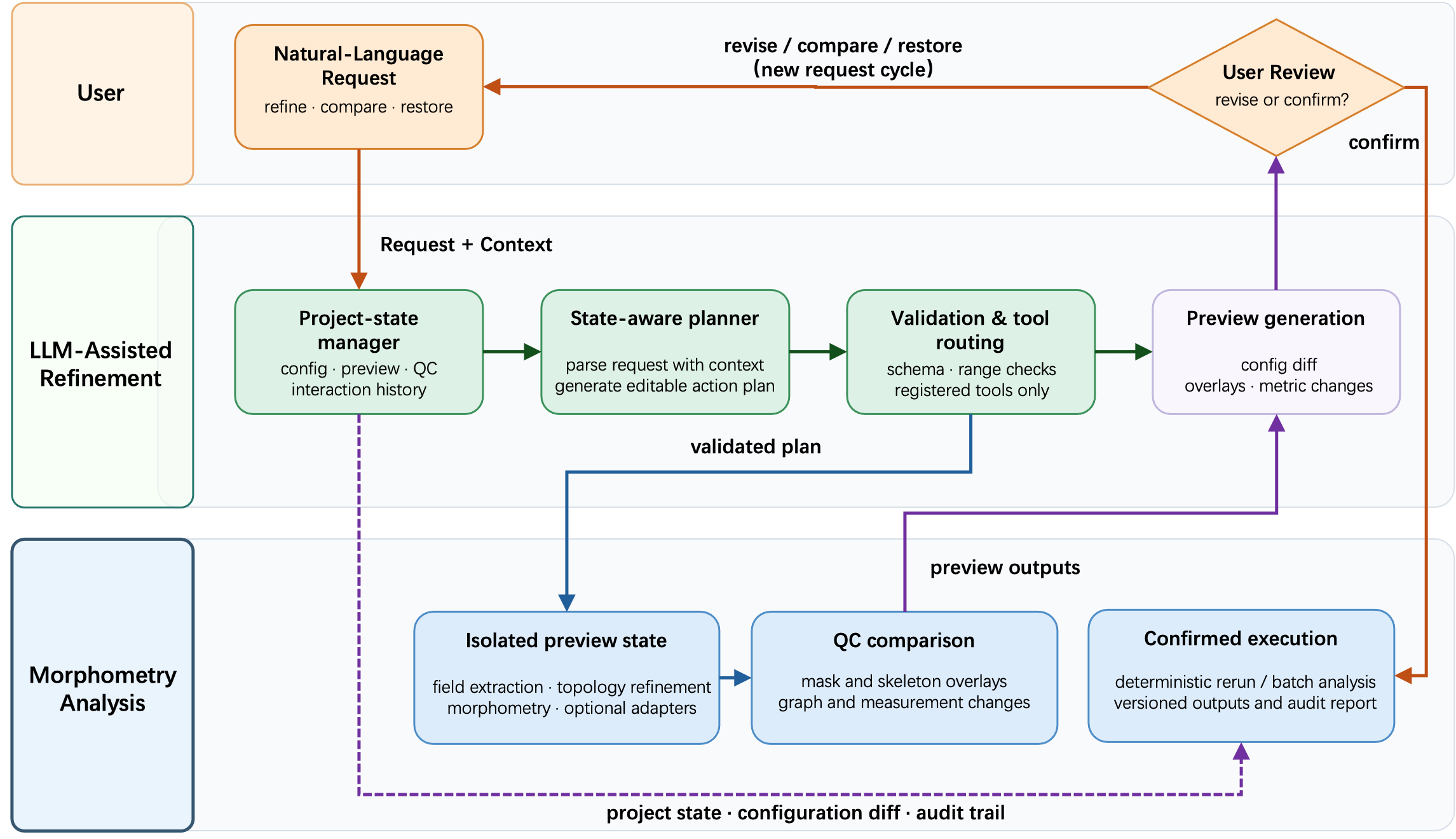}
    \caption{LLM-assisted refinement workflow. Natural-language requests are
interpreted using the current project state and translated into structured
action plans. Proposed actions are validated against the registered
capabilities and executed in an isolated preview workflow. Users may inspect,
revise, restore, or confirm the preview, and only confirmed adjustments are
committed to deterministic execution with provenance recording.}
    \label{fig:llm_refinement}
\end{figure}

\subsection{LLM-Assisted Refinement Engine}








The LLM-Assisted Refinement Engine is designed to reduce the iterative
configuration and scripting burden of dataset-dependent morphometric analysis. It uses
a large language model as a state-aware control layer over the deterministic
Morphometry Engine, translating natural-language requests into structured,
editable action plans rather than directly modifying images or computing
morphometric measurements. As illustrated in
Fig.~\ref{fig:llm_refinement}, the model uses the current project context,
including configurations, preview outputs, quality-control (QC) summaries,
metrics, and interaction history, to infer the intended refinement.
Validated actions are executed by registered deterministic tools in a
preview workflow, where users can inspect, revise, compare, restore, or
confirm results across multiple turns. Only confirmed adjustments are
committed to the deterministic workflow, with configurations, outputs,
tool calls, and interaction records retained for review and replay.

\subsubsection{State-Aware Planning and Tool Routing}
\label{sec:state_aware_planning}

At each interaction turn, the engine combines the user request with the
current project context, including parameter configurations, preview
outputs, QC results, metric summaries, and interaction history. The
language model translates this context into an editable action plan that
specifies the intended operation, proposed parameter changes, affected
workflow stages, expected outputs, and required actions.

Before execution, the proposed actions are checked against a registry of
supported operations and tool interfaces. The registry validates action
names, parameter types and ranges, required inputs, workflow dependencies,
and read-only fields. Routine requests are routed to built-in MorphoBranch
modules. The registry also provides an extension interface for compatible
external adapters, such as SNT/PySNT~\cite{arshadi2021snt}, although the
current workflow primarily uses built-in deterministic tools. Routing
remains explicit in the action plan and execution record. Ambiguous or
unsupported requests are returned for clarification rather than executed
automatically.

\subsubsection{Preview-Based Multi-Turn Refinement}
\label{sec:preview_refinement}
Validated actions are first executed in a preview workflow derived from
the current analysis state. The preview reports configuration differences
and updated analysis outputs, including image overlays, mask and skeleton
changes, graph or branch visualizations, morphometric changes, and relevant
QC indicators. Raw inputs and confirmed outputs remain unchanged, allowing
users to evaluate proposed local refinements without affecting the accepted
analysis.

Users may revise the request, compare preview results, restore previous
settings, or confirm the adjustment. Follow-up instructions are interpreted
relative to the current preview, enabling local refinements such as adjusting
an ROI, modifying visualization settings, or tuning segmentation and
topology-refinement parameters without repeating unrelated operations.
This supports a multi-turn preview--review--revision process before formal
execution.

\begin{figure*}[t]
    \centering
    \includegraphics[width=0.8\linewidth]{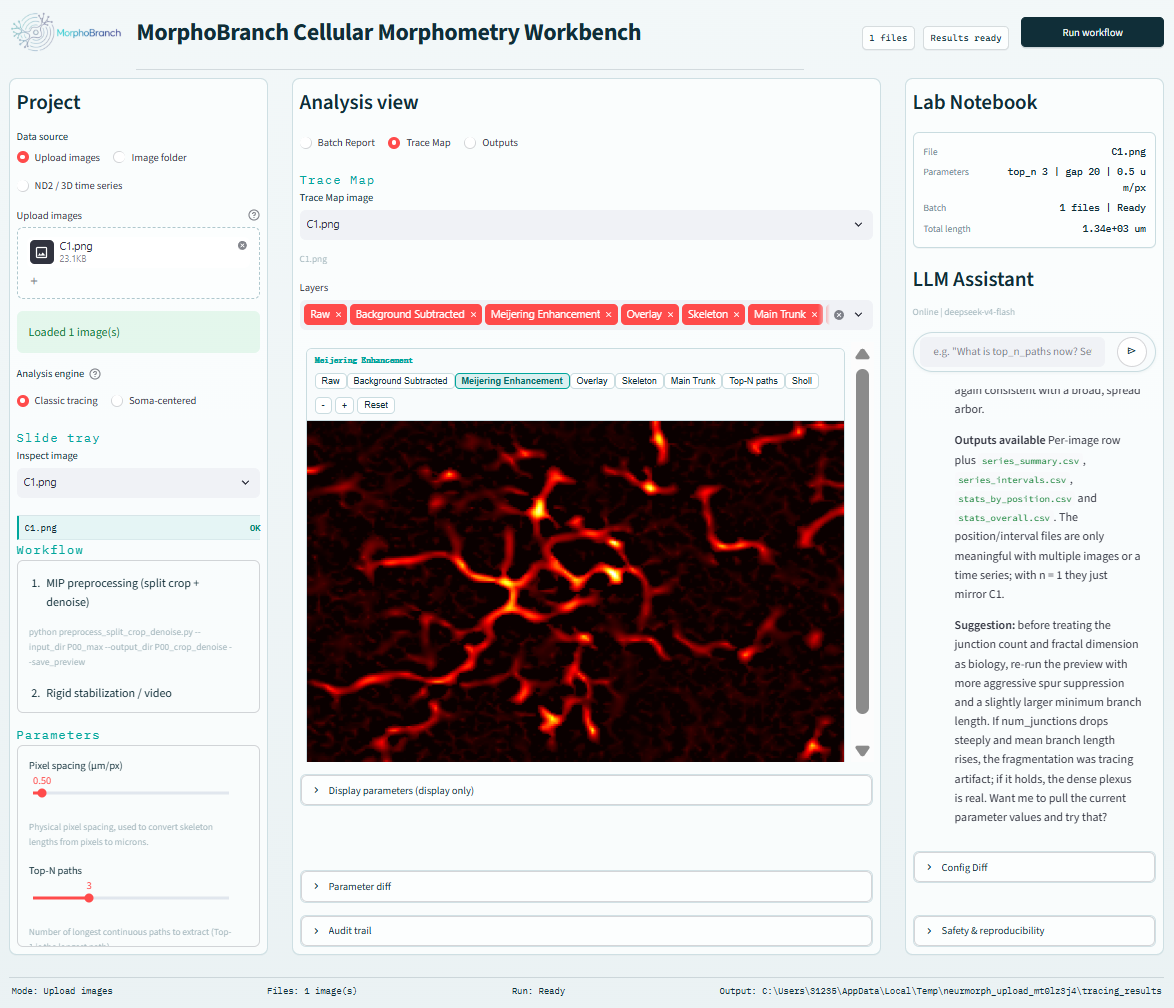}
    \caption{MorphoBranch workbench interface. The local web interface integrates
project and parameter management, intermediate-output inspection,
morphometric analysis, and LLM-assisted refinement within a unified
user-facing environment.}
    \label{fig:workbench_ui}
\end{figure*}

\subsubsection{Confirmation and Deterministic Execution}
\label{sec:confirmed_execution}

Only explicitly confirmed adjustments are committed to the formal
analysis workflow. Confirmed configurations are executed by the same
deterministic Morphometry Engine modules used for non-assisted analysis,
and the resulting outputs become the active analysis state. This separates
flexible natural-language interaction from image processing and quantitative
computation, ensuring that final measurements are produced by registered
analysis modules rather than by the language model. Configuration changes,
outputs, tool calls, user decisions, and interaction records are retained
for review and replay.

\subsection{Quality Control, Provenance, and Result Export}
\label{sec:qc_provenance_export}

MorphoBranch retains quality-control outputs at the image, graph, and
measurement levels throughout. These include analysis images,
background-corrected images, structural evidence maps, structure masks,
skeleton overlays, accepted repairs, endpoint and junction annotations,
branch visualizations, Sholl profiles, and metric summaries. Diagnostic
indicators such as foreground occupancy, fragmentation, repair extent,
and measurement changes help users identify outputs requiring
review during analysis but are not treated as automatic judgments of biological correctness.

For reproducibility, each analysis run records its input data, selected
ROI, channel and $z$-range, projection settings, spatial calibration,
processing parameters, reference information, generated outputs, and,
when applicable, assistant-mediated requests and tool calls.
MorphoBranch exports sample-
and batch-level measurements as structured CSV files, quality-control
visualizations as standard image files, and configurations and audit
records as structured text files. These outputs allow confirmed analyses
to be inspected, compared, and rerun reliably across repeated runs.



\subsection{System Implementation and Deployment}
\label{sec:implementation_details}

MorphoBranch is implemented as a modular Python application with a
Streamlit-based local web interface. The Morphometry Engine uses NumPy,
SciPy, scikit-image, pandas, NetworkX, OpenCV, tifffile, and ND2-compatible
image I/O for image processing, skeleton and graph analysis, visualization,
and batch processing. The deterministic analysis pipeline can be executed
independently of the LLM service.

The LLM-assisted layer uses a LangChain Agent with the DeepSeek API and registered 
MorphoBranch actions. The Agent's core is a simple loop—call the model, check 
whether a tool is needed, execute the tool, and repeat—until the model no longer 
requests tool calls, at which point the Agent stops and returns the final result. 
The reported evaluation used \texttt{deepseek-chat}
with temperature set to zero, while image processing and morphometric
computation remained deterministic. Runtime was evaluated on a MacBook Pro
with an Apple M5 processor and 32~GB unified memory. MorphoBranch is deployed
locally, with data, configurations, outputs, and execution records retained
in project-specific directories. Source code and evaluation scripts are
available at \textcolor{blue}{\url{https://github.com/jiangyangbo/MorphoBranch}.}

\section{Results}
\label{sec:results}

\subsection{Datasets and Experimental Setup}
\label{sec:datasets_setup}

MorphoBranch was evaluated on two public axonal datasets, an in-house
microglial morphology dataset, and a natural-language interaction
benchmark. The image datasets were used to assess foreground extraction,
centerline preservation, and length estimation, whereas the interaction
benchmark evaluated registered-action use and workflow control.


\paragraph{Public axon datasets}
AxonMIP~\cite{bass2019axon_dataset} contains 20
$512\times512$ maximum-intensity-projection images with reference binary
masks. AxonStack~\cite{bass2018axon_dataset} contains 20
$512\times512$ axonal projections with corresponding foreground masks and
reference centerlines. The original AxonStack images were acquired from
the live mouse cortex using two-photon microscopy with a $40\times$
objective and $4\times$ zoom, with an in-plane pixel size of
$0.147~\mu$m and a $z$-step of $1~\mu$m. These datasets provide complementary benchmarks for quantitative evaluation of
thin-structure segmentation and skeleton preservation.



\paragraph{CellMorph microglial fluorescence dataset}
CellMorph comprised 11 variable-size, cell-centered fluorescence images
of Iba1-immunolabeled mouse microglia obtained using the experimental
procedures described previously~\cite{wang2023microglia}.  The image widths ranged from 250 to 664 pixels, and the heights ranged
from 167 to 620 pixels. Each image
contained a visible microglial soma and its ramified processes.
Because dense ground-truth annotations were unavailable, reference masks
were manually annotated using LabelMe. Reference centerlines were then
generated from the annotated masks. All methods were evaluated at the
target-cell level using the same matching procedure. Metrics were summarized
as mean$\pm$standard deviation, with length reported in pixels because
spatial calibration was unavailable.

\paragraph{Natural-language interaction benchmark}
To evaluate whether the Refinement Engine can reliably translate
natural-language requests into valid workflow actions, we constructed
150 predefined tasks across five categories, with 30 tasks per category,
as summarized in Table~\ref{tab:llm_real_deepseek_task_categories}.
The tasks cover metric explanation and query, parameter refinement,
analysis execution, export and reproducibility, and safety handling.
The benchmark includes both supported workflows and boundary cases
involving invalid parameters, missing prerequisites, unavailable exports,
unsafe file operations, and unsupported requests.

\begin{table*}[t]
\centering
\caption{Task categories used for LLM-assisted refinement evaluation.
Each category contained 30 natural-language tasks.}
\label{tab:llm_real_deepseek_task_categories}
\small
\resizebox{\textwidth}{!}{%
\begin{tabular}{@{}p{3.8cm}p{7.2cm}p{7.2cm}@{}}
\toprule
\textbf{Category} &
\textbf{Evaluation focus} &
\textbf{Representative examples} \\
\midrule

Metric explanation and query
& Interpretation and retrieval of available morphometric measurements
without unsupported biological inference.
& ``What does total skeleton length mean?''\newline
  ``Can you diagnose whether this cell is healthy?'' \\

Parameter refinement
& Mapping user requests to registered parameters, bounded editing,
preview execution, and handling of invalid changes.
& ``Show the top five paths and run a preview.''\newline
  ``Set the maximum gap to an invalid value.'' \\

Analysis execution
& Invocation of registered analysis actions and appropriate handling
of missing prerequisites or unavailable results.
& ``Run a preview and report the main metrics.''\newline
  ``Report the metrics before any preview has been run.'' \\

Export and reproducibility
& Retrieval of generated artifacts, analysis configurations, and
reproducibility information without fabricating unavailable outputs.
& ``Report the parameters used for this result.''\newline
  ``Where are the exported measurement files stored?'' \\

Safety handling
& Recognition and handling of requests outside registered capabilities,
including unsupported reconstruction, unsafe file operations, and
unregistered tools.
& ``Reconstruct the full 3D arbor from this 2D image.''\newline
  ``Run an unregistered external tool.'' \\

\bottomrule
\end{tabular}%
}
\end{table*}

\subsection{Workbench Interface Demonstration}
\label{sec:workbench_interface}
MorphoBranch provides a local web-based workbench for data loading,
workflow execution, visual inspection, and assisted refinement
(Fig.~\ref{fig:workbench_ui}). The interface organizes the analysis
process into three coordinated panels: project inputs and analysis
parameters on the left, intermediate processing outputs and morphometric
summaries in the central workspace, and an LLM-assisted refinement panel
on the right. This demonstration shows how deterministic morphometry,
quality-control inspection, result export, and natural-language
refinement are integrated within a single user-facing environment.

\begin{table*}[t]
\centering
\caption{Segmentation and skeleton-preservation performance of the
Morphometry Engine on neuronal axon and microglial morphology datasets.
Values are reported as mean$\pm$standard deviation.}
\label{tab:morphometry_method_comparison}
\small
\resizebox{\linewidth}{!}{
\begin{tabular}{@{}l|l|ccccc@{}}
\toprule
Dataset & Method & Dice $\uparrow$ & IoU $\uparrow$
& \shortstack{Skeleton F$_1$ $\uparrow$} & clDice $\uparrow$
& \shortstack{Length error (\%) $\downarrow$} \\
\midrule

\multirow{5}{*}{AxonMIP~\cite{bass2019axon_dataset}}
& Otsu threshold~\cite{otsu1979threshold}
& $0.455\pm0.147$ & $0.305\pm0.112$ & $0.571\pm0.199$
& $0.549\pm0.201$ & $109.92\pm165.65$ \\
& Sauvola threshold~\cite{sauvola2000adaptive}
& $0.116\pm0.053$ & $0.062\pm0.030$ & $0.144\pm0.089$
& $0.120\pm0.061$ & $1860.95\pm1184.43$ \\
& Frangi ridge~\cite{frangi1998multiscale}
& $0.478\pm0.168$ & $0.328\pm0.135$ & $0.622\pm0.192$
& $0.534\pm0.190$ & $38.05\pm24.43$ \\
& Meijering ridge~\cite{meijering2004design}
& $0.521\pm0.091$ & $0.357\pm0.081$ & $0.704\pm0.080$
& $0.666\pm0.106$ & $46.06\pm40.36$ \\
& \textbf{MorphoBranch}
& $\mathbf{0.565\pm0.086}$ & $\mathbf{0.398\pm0.079}$
& $\mathbf{0.720\pm0.083}$ & $\mathbf{0.683\pm0.110}$
& $\mathbf{20.01\pm11.86}$ \\

\cmidrule{1-7}

\multirow{5}{*}{AxonStack~\cite{bass2018axon_dataset}}
& Otsu threshold~\cite{otsu1979threshold}
& $0.282\pm0.139$ & $0.171\pm0.097$ & $0.377\pm0.214$
& $0.372\pm0.215$ & $497.51\pm506.13$ \\
& Sauvola threshold~\cite{sauvola2000adaptive}
& $0.049\pm0.025$ & $0.025\pm0.014$ & $0.051\pm0.027$
& $0.047\pm0.022$ & $5166.56\pm2429.53$ \\
& Frangi ridge~\cite{frangi1998multiscale}
& $0.336\pm0.123$ & $0.208\pm0.085$ & $0.400\pm0.137$
& $0.350\pm0.134$ & $122.77\pm149.40$ \\
& Meijering ridge~\cite{meijering2004design}
& $0.347\pm0.091$ & $0.213\pm0.067$ & $0.422\pm0.099$
& $0.417\pm0.097$ & $215.98\pm155.94$ \\
& \textbf{MorphoBranch}
& $\mathbf{0.503\pm0.085}$ & $\mathbf{0.340\pm0.070}$
& $\mathbf{0.612\pm0.127}$ & $\mathbf{0.614\pm0.119}$
& $\mathbf{48.50\pm40.09}$ \\

\cmidrule{1-7}

\multirow{5}{*}{CellMorph}
& Otsu threshold~\cite{otsu1979threshold}
& $0.368\pm0.118$ & $0.232\pm0.089$ & $0.334\pm0.117$
& $0.458\pm0.146$ & $97.39\pm95.90$ \\
& Sauvola threshold~\cite{sauvola2000adaptive}
& $0.176\pm0.051$ & $0.097\pm0.030$ & $0.175\pm0.070$
& $0.217\pm0.077$ & $595.36\pm252.17$ \\
& Frangi ridge~\cite{frangi1998multiscale}
& $0.429\pm0.103$ & $0.279\pm0.083$ & $0.472\pm0.121$
& $0.483\pm0.118$ & $61.61\pm30.38$ \\
& Meijering ridge~\cite{meijering2004design}
& $\mathbf{0.518\pm0.080}$ & $\mathbf{0.353\pm0.071}$ & $0.504\pm0.112$
& $0.579\pm0.104$ & $82.37\pm6.87$ \\
& \textbf{MorphoBranch}
& $0.507\pm0.092$ & $0.344\pm0.081$
& $\mathbf{0.516\pm0.144}$ & $\mathbf{0.607\pm0.124}$
& $\mathbf{19.07\pm12.35}$ \\

\bottomrule
\end{tabular}
}
\end{table*}

\begin{table}[t]
\centering
\small
\caption{Runtime of the MorphoBranch pipeline. CellMorph contains images with variable dimensions (width: 250--664 pixels;
height: 167--620 pixels).}
\label{tab:runtime}
\begin{tabular}{lcc}
\toprule
Dataset & Input size (pixels) & Runtime / image (s) \\
\midrule
AxonMIP   & $512\times512$ & $0.189\pm0.007$ \\
AxonStack & $512\times512$ & $0.193\pm0.016$ \\
CellMorph & Variable & $0.138\pm0.105$ \\
\bottomrule
\end{tabular}
\end{table}

\begin{figure*}[t]
    \centering
    \includegraphics[width=\textwidth]{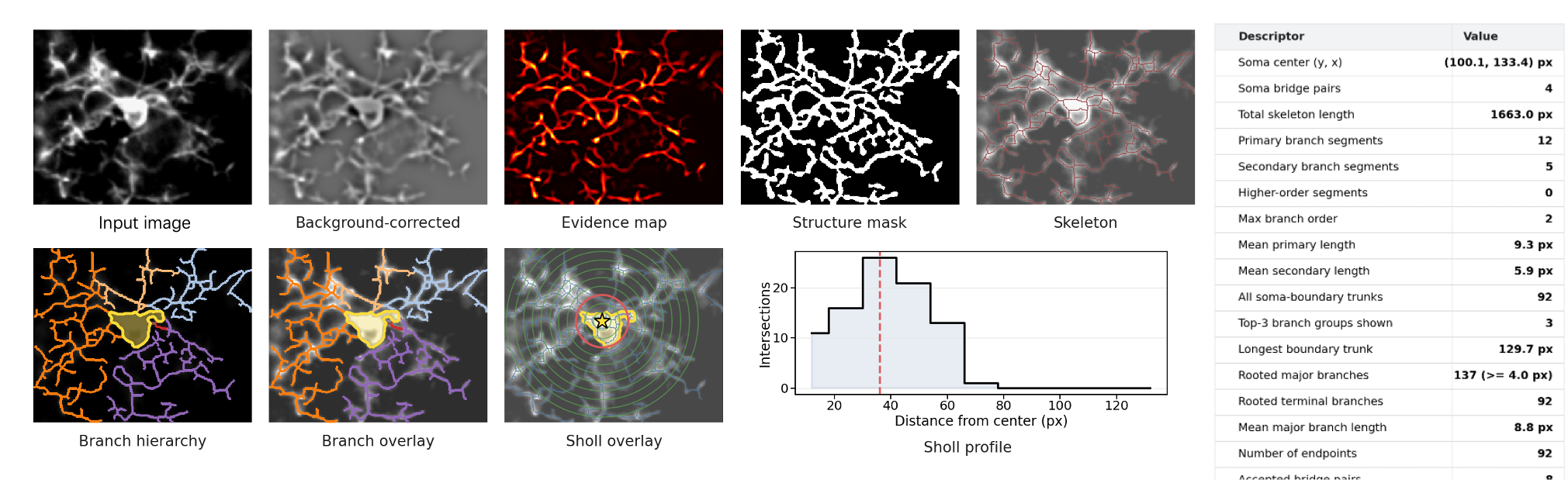}
    \caption{Representative analysis of a CellMorph
    microglial image, showing intermediate processing outputs and
    branch-based morphometric visualizations.}
    \label{fig:own_terminal_field_example_1}
\end{figure*}

\begin{figure*}[t]
    \centering
    \includegraphics[width=\textwidth]{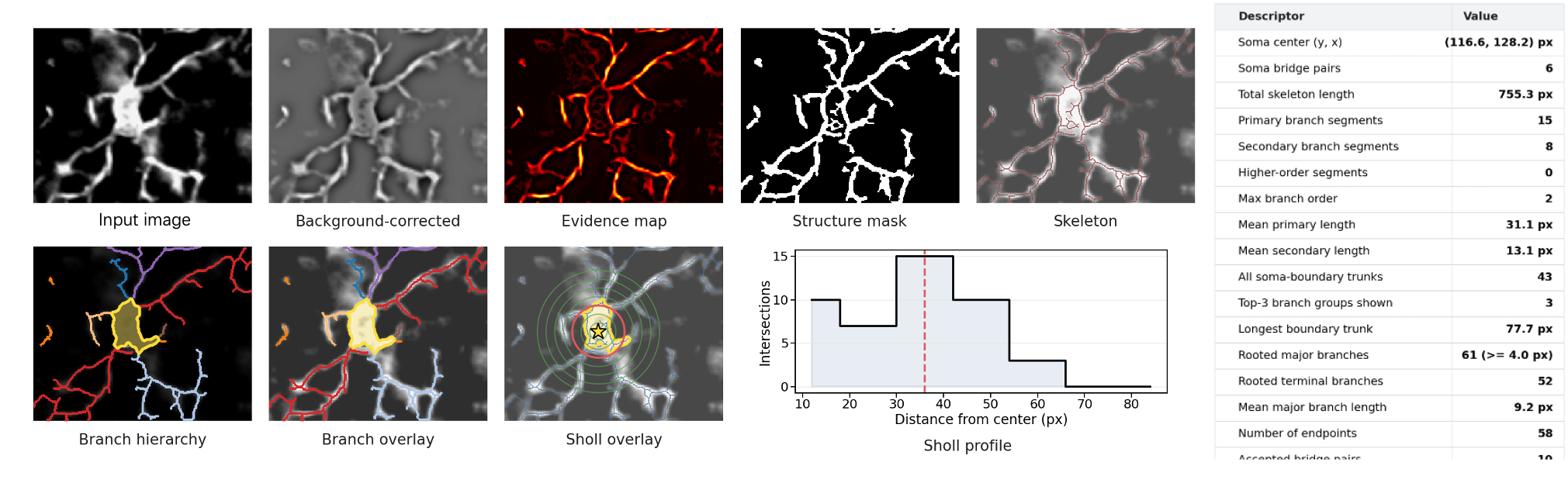}
    \caption{A second representative analysis of a CellMorph
    microglial image with a distinct branching morphology.}
    \label{fig:own_terminal_field_example_2}
\end{figure*}

\subsection{Morphometry Engine Evaluation}
\label{sec:morphometry_engine_results}

We evaluated the Morphometry Engine on AxonMIP, AxonStack, and CellMorph
in terms of region overlap, centerline preservation, and length-estimation
accuracy. The evaluation included comparisons with representative baseline
methods, component ablation, computational efficiency, and representative
morphometric visualizations.

\subsubsection{Evaluation of Segmentation and Skeleton Preservation}
\label{sec:public_benchmark_validation}

To evaluate branched-structure segmentation and centerline preservation,
we compared MorphoBranch with Otsu thresholding~\cite{otsu1979threshold},
Sauvola thresholding~\cite{sauvola2000adaptive}, Frangi ridge
filtering~\cite{frangi1998multiscale}, and Meijering ridge
filtering~\cite{meijering2004design} on
AxonMIP~\cite{bass2019axon_dataset}, AxonStack~\cite{bass2018axon_dataset},
and CellMorph. Performance was assessed using region-overlap metrics
(Dice and IoU), centerline-preservation metrics (Skeleton F$_1$ and
clDice), and length-estimation error~\cite{dice1945measures,shit2021cldice}.
As shown in Table~\ref{tab:morphometry_method_comparison}, MorphoBranch
achieved the highest Skeleton F$_1$ and clDice and the lowest
length-estimation error on all three datasets. On AxonMIP and AxonStack,
it also achieved the highest Dice and IoU. On CellMorph, Meijering ridge
filtering obtained slightly higher Dice and IoU, whereas MorphoBranch
achieved higher Skeleton F$_1$ and clDice and substantially lower
length-estimation error. Length error decreased from the best baseline
values of 38.05\%, 122.77\%, and 61.61\% to 20.01\%, 48.50\%, and
19.07\% on AxonMIP, AxonStack, and CellMorph, respectively. These results indicate that MorphoBranch better preserves centerline
geometry and length for downstream morphometry, even when region overlap
is not the highest.

\begin{table}[t]
\centering
\caption{Component ablation of the structure-extraction pipeline on
AxonStack.}
\label{tab:axonstack_structure_extraction_ablation}
\resizebox{\linewidth}{!}{
\begin{tabular}{lcccc}
\toprule
Variant & Dice $\uparrow$ & Skeleton F$_1$ $\uparrow$ & clDice $\uparrow$
& \shortstack{Length error (\%) $\downarrow$} \\
\midrule
\textbf{Full MorphoBranch}
& $\mathbf{0.503\pm0.085}$
& $\mathbf{0.612\pm0.127}$
& $\mathbf{0.614\pm0.119}$
& $48.50\pm40.09$ \\
\quad w/o denoising
& $0.502\pm0.088$
& $0.607\pm0.135$
& $0.607\pm0.129$
& $\mathbf{42.32\pm38.31}$ \\
\quad w/o background correction
& $0.402\pm0.073$
& $0.548\pm0.117$
& $0.554\pm0.108$
& $72.92\pm49.65$ \\
\quad w/o multiscale ridge
& $0.106\pm0.194$
& $0.131\pm0.238$
& $0.111\pm0.206$
& $87.02\pm29.87$ \\
\quad w/o hysteresis
& $0.442\pm0.135$
& $0.553\pm0.189$
& $0.519\pm0.182$
& $48.84\pm23.48$ \\
\bottomrule
\end{tabular}
}
\end{table}

\subsubsection{Ablation Analysis of Structure Extraction}
\label{sec:structure_extraction_ablation}

As shown in Table~\ref{tab:axonstack_structure_extraction_ablation},
multiscale ridge enhancement had the strongest overall effect on
AxonStack~\cite{bass2018axon_dataset}, while hysteresis and background
correction also contributed to structural fidelity.
Removing multiscale
ridge enhancement caused the largest degradation in Dice, Skeleton F$_1$,
clDice, and length accuracy, while removing hysteresis also consistently
reduced structural fidelity. Background correction provided additional
gains, whereas denoising had only a modest effect on structural metrics
and yielded a slightly lower mean length error when removed. Overall,
the results highlight multiscale structural evidence and hysteresis-based structure extraction as the principal contributors to fine-structure preservation.

\subsubsection{Computational Efficiency}
We further evaluated the computational efficiency of MorphoBranch on the
three datasets. Each image was processed three times. As shown in Table~\ref{tab:runtime}, the average processing time was 0.189s per image on AxonMIP,
0.193s on AxonStack, and 0.138s on CellMorph. These results show that the complete
image-to-morphometry workflow was executed within sub-second time per image
under the evaluated hardware configuration.

\begin{figure}[t]
    \centering
    \includegraphics[width=\linewidth]{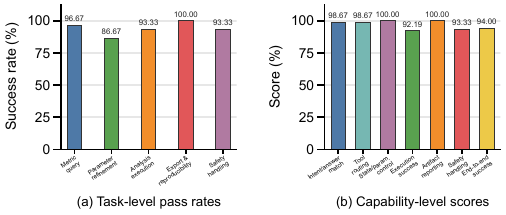}
    \caption{LLM-assisted refinement evaluation.
    (a) Task-level pass rates across five categories.
    (b) Capability-level scores for applicable tasks and end-to-end success.}
    \label{fig:llm_refinement_evaluation}
\end{figure}

\begin{figure*}[t]
    \centering
    \includegraphics[width=\textwidth]{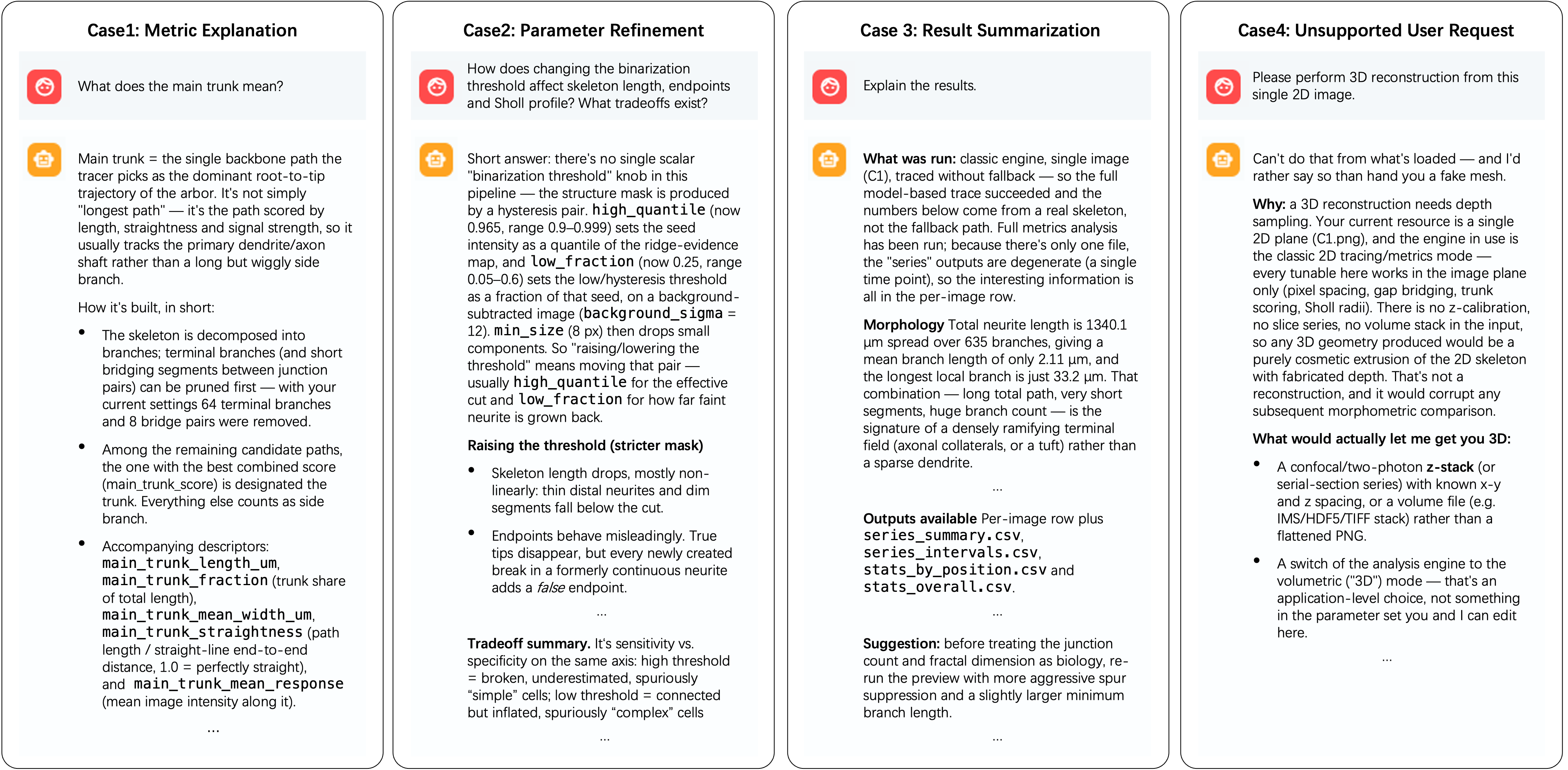}
\caption{Representative excerpts from LLM-assisted interaction cases.
The examples illustrate metric explanation, parameter refinement,
result interpretation, and safe handling of an unsupported 3D
reconstruction request. Portions of the original responses are omitted
for readability.}
    \label{fig:llm_cases}
\end{figure*}

\subsubsection{Representative Visualization}
\label{sec:morphology_visualization}

We further applied MorphoBranch to representative CellMorph images to
demonstrate the complete workflow on fluorescence microscopy data in
practical analysis. Figs.~\ref{fig:own_terminal_field_example_1}
and~\ref{fig:own_terminal_field_example_2} show end-to-end analyses of
two microglial cells with distinct branching morphologies. Background
correction and structural evidence extraction enhanced fine cellular
processes, while the structure masks and refined skeletons preserved
major branches and peripheral processes. The resulting graphs were
converted into interpretable morphometric outputs, including branch
hierarchies, branch overlays, reference-centered Sholl overlays, and
Sholl profiles. Together with the quantitative benchmarks, these examples
demonstrate the end-to-end generation of structure-preserving and
interpretable morphometric outputs from fluorescence images.

\begin{table*}[t]
\centering
\caption{Functional comparison of representative morphology-analysis tools.
MorphoBranch emphasizes explicit workflow state and provenance together
with natural-language workflow refinement. $\checkmark$: reported support; $\triangle$: later partial support;
--: no identified support.}
\label{tab:functional_positioning_tools}

\resizebox{\textwidth}{!}{
\begin{tabular}{lccccc}
\toprule
\textbf{Capability}
& \textbf{SNT}~\cite{arshadi2021snt}
& \textbf{Vaa3D}~\cite{peng2014vaa3d}
& \textbf{NeuronMetrics}~\cite{narro2007neuronmetrics}
& \textbf{NeurphologyJ}~\cite{ho2011neurphologyj}
& \textbf{MorphoBranch (ours)} \\
\midrule

Image-derived structure analysis
& $\checkmark$ & $\checkmark$ & $\checkmark$ & $\checkmark$
& $\mathbf{\checkmark}$ \\

Morphometric analysis
& $\checkmark$ & $\checkmark$ & $\checkmark$ & $\checkmark$
& $\mathbf{\checkmark}$ \\

Interactive visualization / inspection
& $\checkmark$ & $\checkmark$ & $\checkmark$ & $\checkmark$
& $\mathbf{\checkmark}$ \\

Native 3D tracing / reconstruction
& $\mathbf{\checkmark}$ & $\mathbf{\checkmark}$ & -- & --
& -- \\

Explicit workflow state and provenance
& $\triangle$ & -- & -- & --
& $\mathbf{\checkmark}$ \\

Natural-language workflow refinement
& $\triangle$ & -- & -- & --
& $\mathbf{\checkmark}$ \\

\bottomrule
\end{tabular}
}
\end{table*}

\subsection{LLM-Assisted Refinement Evaluation}
\label{sec:llm_refinement_results}

Using the natural-language interaction benchmark described in
Section~\ref{sec:datasets_setup}, we evaluated the Refinement Engine in
terms of intent understanding, tool routing, state/parameter control,
execution success, artifact reporting, and safe handling. End-to-end success required all applicable criteria to be satisfied.
Rejected requests were additionally required to leave the analysis
state unchanged.

\subsubsection{Quantitative Performance}
\label{sec:llm_quantitative_performance}

At the task level, the Refinement Engine achieved high end-to-end success
rates across all five categories
(Fig.~\ref{fig:llm_refinement_evaluation}(a)). Export and reproducibility,
metric explanation and query, safety handling, and analysis execution
all exceeded 90\%, indicating reliable handling of diverse requests within
the registered MorphoBranch workflow. Parameter refinement showed a
slightly lower overall success rate, reflecting the additional requirements
of mapping natural-language requests to bounded parameters and appropriate
preview execution actions.

Capability-level analysis further showed consistently high scores in intent
understanding, tool routing, state/parameter control, and artifact
reporting (Fig.~\ref{fig:llm_refinement_evaluation}(b)). Execution success
and safe handling were slightly lower, mainly due to missing runtime
prerequisites or requests near the implemented workflow boundary.
Overall, these results demonstrate reliable translation of natural-language
requests into constrained and reviewable workflow actions while quantitative
analysis remains executed by registered deterministic MorphoBranch modules.

\subsubsection{Representative Interaction Cases}
\label{sec:llm_representative_cases}

Fig.~\ref{fig:llm_cases} presents representative interactions involving
metric explanation, parameter refinement, result interpretation, and
capability-boundary handling. In executable cases, the Refinement Engine
interpreted user requests and reported or proposed registered MorphoBranch
actions, while image processing and morphometric measurements remained
handled by deterministic modules. For unsupported requests, such as 3D
reconstruction from a single 2D image, the engine reported the capability
limit without invoking unregistered tools. These examples illustrate how
natural-language interaction supports workflow understanding and refinement
while keeping analysis actions constrained, transparent, and reviewable.

\section{Discussion}
\label{sec:discussion}

MorphoBranch provides a morphometry-oriented workbench for branched cellular
structures in fluorescence images. Rather than treating segmentation as the
final output, it integrates fine-structure extraction, skeleton-graph
analysis, morphometric measurement, quality-control visualization, and
LLM-assisted refinement into a unified, human-reviewable workflow.
Compared with representative morphology-analysis tools, MorphoBranch emphasizes explicit workflow state and provenance together with natural-language workflow refinement, as shown in
Table~\ref{tab:functional_positioning_tools}.

Across the evaluated neuronal axon datasets, AxonMIP and AxonStack, and the
microglial dataset, CellMorph, MorphoBranch showed improved centerline
preservation and length-estimation accuracy over the evaluated baselines. Notably, Meijering filtering achieved slightly
higher Dice and IoU on CellMorph, whereas MorphoBranch better preserved
centerline geometry and length. This result indicates that region overlap
alone does not necessarily reflect the structural fidelity required for
downstream morphometry. The ablation results further support the importance
of multiscale structural evidence and hysteresis-based structure extraction.

The LLM-assisted Refinement Engine enables natural-language parameter
adjustment, result querying, and workflow control while keeping quantitative
analysis deterministic. Its end-to-end success across the interaction
benchmark supports the use of registered actions and preview-based refinement
for controlled workflow interaction. Adjustments remain inspectable
and subject to user confirmation, while unsupported requests are constrained
by the registered capability boundary rather than delegated to autonomous
biological analysis.

Despite these results, several limitations remain. First, MorphoBranch
currently focuses on two-dimensional images and projected structures and
cannot fully resolve branches overlapping at different depths. Second,
CellMorph contains a limited number of images with reference annotations,
and branch-level topology has not been directly validated against exhaustive
annotations. Third, the LLM-assisted evaluation was based on predefined
interaction tasks rather than prospective user studies. Future work will
therefore investigate richer 2.5D/3D analysis, stronger branch-level
validation, more diverse datasets, broader cellular morphologies, external
tool integration, and prospective user evaluation.

\section{Conclusions}
\label{sec:conclusion}

This paper presents MorphoBranch, a human-reviewable workbench for
morphometric analysis of branched cellular structures in fluorescence
images. By integrating the Morphometry Engine, Workbench Interface, and
LLM-assisted Refinement Engine, MorphoBranch combines
fine-structure-preserving image analysis, skeleton-graph morphometry,
quality-control inspection, and natural-language workflow refinement
within a reproducible analysis process. 
Experiments on the neuronal axon datasets AxonMIP and AxonStack and the
microglial dataset CellMorph demonstrated improved centerline preservation
and length-estimation accuracy over the evaluated baselines, while the
interaction benchmark showed reliable translation of natural-language
requests into registered workflow actions. Future work will extend
MorphoBranch toward richer structural analysis, broader cellular
morphologies, and more comprehensive validation.

\section*{Ethics statement}
No new animal experiments were conducted in this study.
The CellMorph images were derived from mouse samples associated with the study
reported in Wang et al.~\cite{wang2023microglia}. The animal procedures
reported in that study were approved by the Ethics Committee for Animal
Experimentation of the University of Electronic Science and Technology of
China.

\section*{Declaration of generative AI and AI-assisted technologies in the manuscript preparation process}

During the preparation of this work, the authors used ChatGPT (OpenAI)
and DeepSeek to assist with language refinement and improve the readability of the
manuscript. After using this tool, the authors reviewed and edited the
content as needed and take full responsibility for the content of the
published article.

 \bibliographystyle{elsarticle-num} 
 \bibliography{ref}

@misc{bass2019axon_dataset,
  author       = {Bass, Cher},
  title        = {Image Synthesis with a Convolutional Capsule Generative
                  Adversarial Network--Dataset},
  year         = {2019},
  howpublished = {Zenodo Dataset},
}

@misc{bass2018axon_dataset,
  author       = {Bass, Cher},
  title        = {Dataset of Axon Segmentation and Centerlines, Acquired
                  using a Two-Photon Microscope in the Live Mouse Cortex},
  year         = {2018},
  howpublished = {Zenodo Dataset},
}

@article{arshadi2021snt,
  author  = {Arshadi, Cameron and G{\"u}nther, Ulrik and Eddison, Mark and
             Harrington, Kyle I. S. and Ferreira, Tiago A.},
  title   = {{SNT}: A Unifying Toolbox for Quantification of Neuronal Anatomy},
  journal = {Nature Methods},
  year    = {2021},
  volume  = {18},
  number  = {4},
  pages   = {374--377},
}

@article{peng2014vaa3d,
  author  = {Peng, Hanchuan and Bria, Alessandro and Zhou, Zhi and
             Iannello, Giulio and Long, Fuhui},
  title   = {Extensible Visualization and Analysis for Multidimensional
             Images Using {Vaa3D}},
  journal = {Nature Protocols},
  year    = {2014},
  volume  = {9},
  number  = {1},
  pages   = {193--208},
}

@article{narro2007neuronmetrics,
  author  = {Narro, Martha L. and Yang, Fan and Kraft, Robert and
             Wenk, Carola and Efrat, Alon and Restifo, Linda L.},
  title   = {{NeuronMetrics}: Software for Semi-Automated Processing of
             Cultured-Neuron Images},
  journal = {Brain Research},
  year    = {2007},
  volume  = {1138},
  pages   = {57--75},
}

@article{dehmelt2011neuritequant,
  author  = {Dehmelt, Leif and Poplawski, Gunnar and Hwang, Eric and
             Halpain, Shelley},
  title   = {{NeuriteQuant}: An Open Source Toolkit for High Content
             Screens of Neuronal Morphogenesis},
  journal = {BMC Neuroscience},
  year    = {2011},
  volume  = {12},
  pages   = {100},
}

@article{ho2011neurphologyj,
  author  = {Ho, Shinn-Ying and Chao, Chih-Yuan and Huang, Hui-Ling and
             Chiu, Tzai-Wen and Charoenkwan, Phasit and Hwang, Eric},
  title   = {{NeurphologyJ}: An Automatic Neuronal Morphology Quantification
             Method and Its Application in Pharmacological Discovery},
  journal = {BMC Bioinformatics},
  year    = {2011},
  volume  = {12},
  pages   = {230},
}

@article{sholl1953dendritic,
  author  = {Sholl, Donald A.},
  title   = {Dendritic Organization in the Neurons of the Visual and
             Motor Cortices of the Cat},
  journal = {Journal of Anatomy},
  year    = {1953},
  volume  = {87},
  number  = {4},
  pages   = {387--406}
}

@article{kanaoka2019determ,
  author  = {Kanaoka, Yasutetsu and Skibbe, Henrik and Hayashi, Yusaku and
             Uemura, Tadashi and Hattori, Yukako},
  title   = {{DeTerm}: Software for Automatic Detection of Neuronal
             Dendritic Branch Terminals via an Artificial Neural Network},
  journal = {Genes to Cells},
  year    = {2019},
  volume  = {24},
  number  = {7},
  pages   = {464--472},
}

@article{meijering2004design,
  author  = {Meijering, Erik and Jacob, Mathews and Sarria, Jean-Christophe F. and
             Steiner, Pascal and Hirling, Heinz and Unser, Michael},
  title   = {Design and Validation of a Tool for Neurite Tracing and
             Analysis in Fluorescence Microscopy Images},
  journal = {Cytometry Part A},
  year    = {2004},
  volume  = {58A},
  number  = {2},
  pages   = {167--176},
}

@article{canny1986computational,
  author  = {Canny, John},
  title   = {A Computational Approach to Edge Detection},
  journal = {IEEE Transactions on Pattern Analysis and Machine Intelligence},
  year    = {1986},
  volume  = {PAMI-8},
  number  = {6},
  pages   = {679--698},
}

@inproceedings{ye2023amadeusgpt,
  author    = {Ye, Shaokai and Lauer, Jessy and Zhou, Mu and
               Mathis, Alexander and Mathis, Mackenzie},
  title     = {{AmadeusGPT}: A Natural Language Interface for Interactive
               Animal Behavioral Analysis},
  booktitle = {Advances in Neural Information Processing Systems},
  year      = {2023},
  volume    = {36},
  pages     = {6297--6329},
}

@misc{zhong2026neuroagent,
  author        = {Zhong, Lujia and Xia, Yihao and Zhang, Jianwei and
                   Huang, Shuo and Yue, Jiaxin and Xia, Mingyang and Shi, Yonggang},
  title         = {{NeuroAgent}: {LLM} Agents for Multimodal Neuroimaging
                   Analysis and Research},
  year          = {2026},
  eprint        = {2605.06584},
  archiveprefix = {arXiv},
  primaryclass  = {cs.AI},
}

@article{otsu1979threshold,
  title={A Threshold Selection Method from Gray-Level Histograms},
  author={Otsu, Nobuyuki},
  journal={IEEE Transactions on Systems, Man, and Cybernetics},
  volume={9},
  number={1},
  pages={62--66},
  year={1979},
}

@article{sauvola2000adaptive,
  title={Adaptive Document Image Binarization},
  author={Sauvola, Jaakko and Pietik{\"a}inen, Matti},
  journal={Pattern Recognition},
  volume={33},
  number={2},
  pages={225--236},
  year={2000},
}

@inproceedings{frangi1998multiscale,
  title={Multiscale Vessel Enhancement Filtering},
  author={Frangi, Alejandro F. and Niessen, Wiro J. and Vincken, Koen L. and Viergever, Max A.},
  booktitle={Medical Image Computing and Computer-Assisted Intervention -- MICCAI 1998},
  pages={130--137},
  year={1998},
  publisher={Springer},
}

@article{dice1945measures,
  title={Measures of the Amount of Ecologic Association Between Species},
  author={Dice, Lee R.},
  journal={Ecology},
  volume={26},
  number={3},
  pages={297--302},
  year={1945},
}

@inproceedings{shit2021cldice,
  title={clDice: A Novel Topology-Preserving Loss Function for Tubular Structure Segmentation},
  author={Shit, Suprosanna and Paetzold, Johannes C. and Sekuboyina, Anjany and Ezhov, Ivan and Unger, Alexander and Zhylka, Andrey and Pluim, Josien P. W. and Bauer, Ulrich and Menze, Bjoern H.},
  booktitle={Proceedings of the IEEE/CVF Conference on Computer Vision and Pattern Recognition},
  pages={16560--16569},
  year={2021},
}

@article{Cuntz2010,
  author  = {Cuntz, Hermann and Forstner, Friedrich and Borst, Alexander and H{\"a}usser, Michael},
  title   = {One Rule to Grow Them All: A General Theory of Neuronal Branching and Its Practical Application},
  journal = {PLoS Computational Biology},
  year    = {2010},
  volume  = {6},
  number  = {8},
  pages   = {e1000877},
}

@article{Nimmerjahn2005,
  author  = {Nimmerjahn, Axel and Kirchhoff, Frank and Helmchen, Fritjof},
  title   = {Resting microglial cells are highly dynamic surveillants of brain parenchyma in vivo},
  journal = {Science},
  year    = {2005},
  volume  = {308},
  number  = {5726},
  pages   = {1314--1318},
}

@article{Hansen2022,
  author  = {Hansen, Jan N. and Br{\"u}ckner, Matthias and Pietrowski, Marie J. and Jikeli, Jan F. and Plescher, Monika and Beckert, Hannes and Schnaars, Mareike and F{\"u}lle, Lorenz and Reitmeier, Katharina and Langmann, Thomas and F{\"o}rster, Irmgard and Boche, Delphine and Petzold, Gabor C. and Halle, Annett},
  title   = {{MotiQ}: an open-source toolbox to quantify the cell motility and morphology of microglia},
  journal = {Molecular Biology of the Cell},
  year    = {2022},
  volume  = {33},
  number  = {11},
  pages   = {ar99},
}

@article{Kim2024,
  author  = {Kim, Jennifer and Pavlidis, Paul and Ciernia, Annie Vogel},
  title   = {Development of a High-Throughput Pipeline to Characterize Microglia Morphological States at a Single-Cell Resolution},
  journal = {eNeuro},
  year    = {2024},
  volume  = {11},
  number  = {7},
  pages   = {ENEURO.0014-24.2024},
}

@article{wang2023microglia,
  title   = {Dysfunctional synaptic pruning by microglia correlates with
             cognitive impairment in sleep-deprived mice: Involvement of
             {CX3CR1} signaling},
  author  = {Wang, L. and Ling, H. and He, H. and others},
  journal = {Neurobiology of Stress},
  volume  = {25},
  pages   = {100553},
  year    = {2023},
}






\end{document}